# High-throughput screening for boride superconductors

Shiya Chen[1], Zepeng Wu[1], Zhen Zhang[2], Shunqing Wu[1], Kai-Ming Ho[2], Vladimir Antropov[3] and Yang Sun[1*]

[1]Department of Physics, Xiamen University, Xiamen 361005, China
[2]Department of Physics, Iowa State University, Ames, Iowa 50011, USA
[3]Ames National Laboratory, Ames, Iowa 50011, USA

**Abstract**

A high-throughput screening using density functional calculations is performed to search for stable boride superconductors from the existing materials database. The workflow employs the fast frozen phonon method as the descriptor to evaluate the superconducting properties quickly. 23 stable candidates were identified during the screening. The superconductivity was obtained earlier experimentally or computationally for almost all found binary compounds. Previous studies on ternary borides are very limited. Our extensive search among ternary systems confirmed superconductivity in known systems and found several new compounds. Among these discovered superconducting ternary borides, $TaMo_2B_2$ shows the highest superconducting temperature of ~12K. Most predicted compounds were synthesized previously; therefore, our predictions can be examined experimentally. Our work also demonstrates that the boride systems can have diverse structural motifs that lead to superconductivity.

[*]Email: yangsun@xmu.edu.cn

## I. Introduction

Superconducting materials have numerous applications in the modern society. The search for new superconductors with high critical superconducting temperature ($T_c$) is key to many future techniques in energy, medical care, transportation, and quantum computing. However, the conventional approaches to discovering new superconducting materials through direct experimental synthesis are time-consuming and resource-intensive. In recent years, it has been shown that computational prediction and design can greatly facilitate the discovery of new superconducting materials[1,2]. The computational algorithms based on the density-functional perturbation theory (DFPT) can provide a satisfactory description of the electron-phonon coupling (EPC) and $T_c$ for conventional superconductors. Due to the significant cost of the DFPT calculations, $T_c$ calculations are now combined with information technologies such as data mining, machine learning, and high-throughput screening to guide the theoretical search of conventional superconductors[3–6].

The discovery of superconductivity in the structurally simple $MgB_2$ compound[6,7] has stimulated substantial research efforts toward uncovering phonon-mediated superconductors within similar chemistries. Many attempts have been made to identify new superconducting phases derived from the $MgB_2$ compound, such as chemical doping or substitution, resulting in compounds like $Mg_{1-x}Li_xB_2$[8], and $Mg_{1-x}Zr_xB_2$[9], and $MgB_2$-like boride systems such as $MoB_2$[10] and $WB_2$[11]. Recently, we have shown that the Brillion zone-centered EPC strength can be a simple descriptor to identify phonon-mediated superconducting materials in hydrides and borides[12–17]. This greatly reduced the computational cost in high-throughput screening of boride superconductors. While we identified a few novel boride systems that exhibit an interesting high $T_c$, their thermodynamic stabilities are not sufficiently high for synthesis, and the structural motifs are still very close to $MgB_2$[13]. Therefore, it remains unclear if other stable structural motifs in the borides can show feasible superconducting properties. In this study, we perform high-throughput screening on binary or ternary boride superconductors under ambient pressure from an existing materials database. These materials are stable and mostly synthesized before. We will use the zone-centered EPC strength as the main descriptor to identify their possible superconducting behavior. The structural motifs that can lead to possible $T_c$ will be analyzed for the boride systems.

## II. Methods

Crystal structures are optimized by density functional theory (DFT) calculations, performed using the projector augmented wave (PAW) method[18] within DFT as implemented in the VASP code[19,20]. The exchange and correlation energy is treated with generalized gradient approximation (GGA) and parameterized by the Perdew-Burke-Ernzerhof formula (PBE)[21]. A plane-wave basis was used with a kinetic energy cutoff of 520 eV, and the convergence criterion for the total energy was set to $10^{-5}$ eV. Monkhorst−Pack's sampling scheme[22] was adopted for Brillouin zone sampling with a $\mathbf{k}$-point grid of $2\pi \times 0.033$ Å$^{-1}$. The lattice vectors (supercell shape and size) and atomic coordinates are fully relaxed until the force on each atom is less than 0.01 eV/Å. The initial screening of crystal structure was based on the MPRester package, which allows for the materials screening from the Materials Project database[23].

The high-throughput screening of strong EPC in these metal borides is based on the fast frozen-phonon calculation of the zone-center EPC strength[12] defined by

$$\lambda_\Gamma = \sum_v \lambda_{\Gamma v}, \qquad (1)$$

where $\sum_v$ indicates the summation of all modes at zone-center $\Gamma$. $\lambda_{\Gamma v}$ is defined by

$$\lambda_{\Gamma v} = \frac{\widetilde{\omega}_{\Gamma v}^2 - \omega_{\Gamma v}^2}{4\omega_{\Gamma v}^2}, \qquad (2)$$

where $\widetilde{\omega}_{\Gamma v}$ and $\omega_{\Gamma v}$ are unscreened and screened phonon frequencies of mode $v$ at the zone-center, respectively. The screened phonon frequencies were calculated using the primitive cell and finite displacement methods implemented in the Phonopy code[24]. The displacement amplitude in the frozen-phonon calculations is 0.02 Å. The convergence criterion of the total energy is $10^{-8}$ eV.

The calculations of full Brillouin-zone EPC constants and the superconducting temperature $T_c$ were performed based on density functional perturbation theory (DFPT) [25] implemented in the Quantum ESPRESSO code[19,26,27]. Ultrasoft pseudopotentials[28] with PBE functional were used with a kinetic energy cutoff of 80 Ry and a charge density cutoff of 800 Ry. After the convergence test, the plane-wave cutoff and the charge density cutoff were chosen to be 80 and 640 Ry, respectively. Monkhorst−Pack's sampling scheme[22] was adopted for Brillouin-zone sampling with a $\mathbf{k}$-point grid of $2\pi \times 0.025$ Å$^{-1}$. Self-consistent field (SCF) calculations were performed with a dense $\mathbf{k}$ mesh of twice the scale of the sampling scheme, followed by the DFPT calculation with the $\mathbf{k}$ mesh of the same sampling scheme and set the $\mathbf{q}$ mesh to half of the $\mathbf{k}$ mesh. We used 0.015 Ry as the Gaussian broadening for $T_c$ calculation. The test on TaMo$_2$B$_2$ in Fig. S18 suggests the setting is sufficient to converge the $T_c$ in 1-2 K.

The isotropic Eliashberg spectral function was obtained via the average over the Brillouin

zone[29,30],

$$\alpha^2(\omega)F(\omega) = \frac{1}{2N(\epsilon_F)}\sum_{qv}\frac{\gamma_{qv}}{\hbar\omega_{qv}}\delta(\omega-\omega_{qv}), \quad (3)$$

where $N(\epsilon_F)$ is the density of states at the Fermi level $\epsilon_F$; $\omega_{qv}$ denotes the phonon frequency of mode ν with wave vector q. $\gamma_{qv}$ is the phonon linewidth defined by

$$\gamma_{qv} = \frac{2\pi\omega_{qv}}{\Omega_{BZ}}\sum_{ij}\int d^3k \left|g_{k,qv}^{ij}\right|^2 \delta(\epsilon_{q,i}-\epsilon_F)\delta(\epsilon_{k+q,j}-\epsilon_F), \quad (4)$$

where $g_{k,qv}^{ij}$ is the EPC matrix element; $\epsilon_{q,i}$ and $\epsilon_{k+q,j}$ are eigenvalues of Kohn-Sham orbitals at bands i, j, and wave vectors $\boldsymbol{q}$, $\boldsymbol{k}$. The full Brillouin zone EPC constant λ is determined through the integration of the Eliashberg spectral function,

$$\lambda = 2\int\frac{\alpha^2(\omega)F(\omega)}{\omega}d\omega. \quad (5)$$

The $T_c$ is obtained with the analytical McMillan formula[29] modified by Allen−Dynes[15,31]

$$T_c = \frac{\omega_{\log}}{1.2}exp\left[\frac{-1.04(1+\lambda)}{\lambda(1-0.62\mu^*)-\mu^*}\right], \quad (6)$$

where $\omega_{\log}$ is the logarithmic average frequency

$$\omega_{\log} = \exp\left[\frac{2}{\lambda}\int\frac{d\omega}{\omega}\alpha^2(\omega)F(\omega)\ln\omega\right], \quad (7)$$

and $\mu^*$ is the effectively screened Coulomb repulsion constant, set as 0.1 in our calculations.

### III. Results and discussions

3.1 High throughput screening

Figure 1 summarizes the screening strategy for superconducting boride compounds[23]. We design five criteria to filter low-energy binary or ternary borides for promising superconducting materials from the structure database[32]. These criteria include a chemical composition filter (C), a magnetic filter (M), a stability filter (S), a bands gap filter (B), and an EPC strength filter (E). These filters are defined as follows.

The C-filter selects binary or ternary boride compounds containing fewer than 40 atoms in the primitive cell. These compounds contain various elements, as shown in Fig. 1(b). We eliminated compounds with *4f* and *5f* elements except for La due to the problematic DFT calculation for *f* electrons. The primary goal is to pinpoint simple binary or ternary borides for further consideration. This step effectively narrows down the candidate pool, resulting in 1678 phases for further evaluation.

The M-filter removes compounds with finite magnetic moment (larger than 0.01 μ$_B$/atom)

from the pool because the superconducting and magnetic phases are mutually exclusive.

The S-filter attempts to select stable or low-energy metastable phases that exist in experiments or have a large chance of being synthesized. We employ the energy above the convex hull ($E_d$) less than 0.1 eV/atom as the criterion. $E_d$ measures the energy of a material to decompose into a set of more stable compounds. Larger $E_d$ indicates poorer stability, while a zero $E_d$ indicates the most stable phase. 0.1 eV/atom corresponds to the energy of thermal fluctuation ~1100 K, a typical energy scale for the boride phase that can be synthesized experimentally[15]. The filter removed 377 unstable phases from the pool.

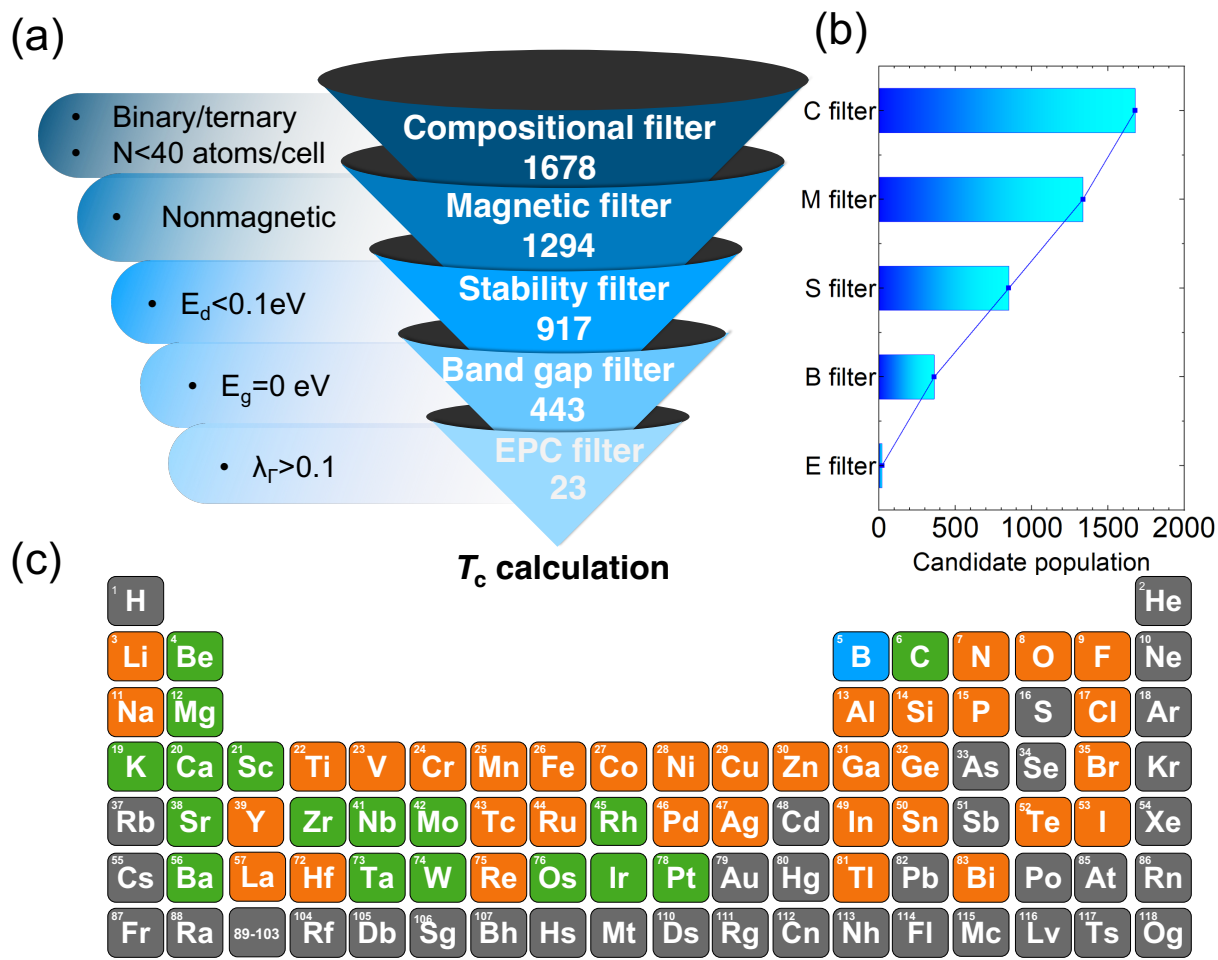

**Fig. 1** (a) Schematics of screening workflow. (b) The bar chart of populations after each filter. (c) The elements involved in the screening. Orange indicates the elements in the initial pool after the B filter. Green indicates the remaining elements after the E filter.

The B-filter selects metallic materials with zero band gap, a characteristic commonly associated with superconductors. By implementing this filter, we eliminate semiconductors or insulators that, due to their electronic structure, are unlikely to exhibit superconducting behavior.

By applying these filters and quantifying the number of materials at each step, we ensure the selected compounds align with our design principle for potential superconductivity. Figure 1(a) shows the screening results in 443 candidates after B-filter. Before the final $T_c$ calculations, we apply the E-filter to identify structures with strong EPC strength based on the fast frozen-phonon calculations of zone-center EPC strength[12]. This method is particularly efficient in identifying phonon-mediated superconductors in the borides and hydrides, where the zone-center phonon modes contributed significantly to the EPC[13,15,16,33]. By setting a threshold of 0.1 for the zone-center EPC strength ($\lambda_\Gamma > 0.1$), we ultimately identified 23 materials for further superconductivity calculations. The zone-center EPC contributions are shown in Fig. S1-11. Comparing the candidate population after each filter in Fig. 1(b), it is evident that the EPC screening reduced the most significant fraction (95%) of the parent structure pool, suggesting strong EPC as a rare property in the materials.

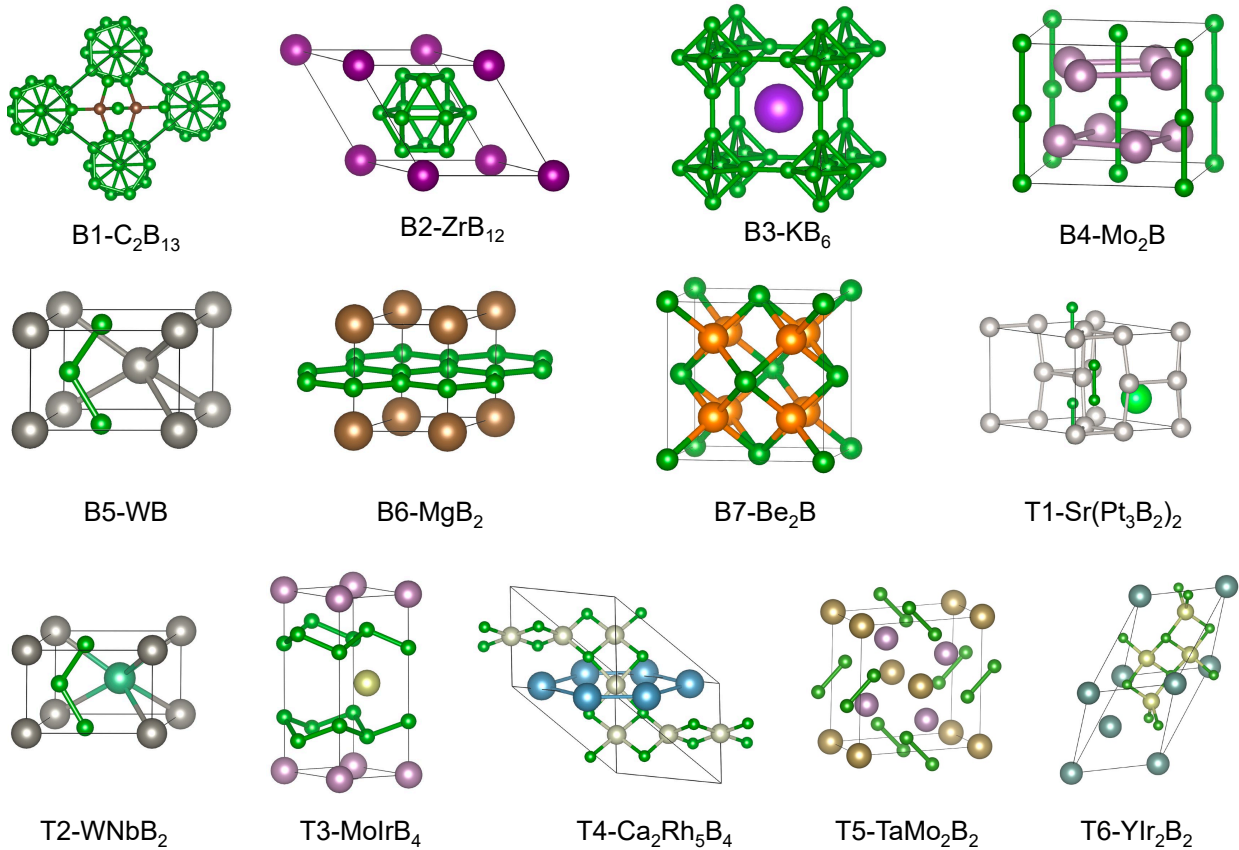

**Fig. 2** Structural Motifs of potential superconducting boride phases.

Table 1 The structural and superconductivity properties of the screened candidate. $\lambda_\Gamma$ is from fast zone-center EPC calculations, while $\lambda_{BZ}$ is EPC from full Brillouin zone calculation. Star ($^*$) indicates the calculated result from this work. "Y" indicates the compound was previously synthesized according to the Inorganic Crystal Structure Database (ICSD)[34]. "N" indicates the compound was not synthesized or non-superconducting in experiments. "?" indicates the lack of experimental investigation on superconductivity. "-" indicates the $T_c$ was not calculated theoretically.

| Structural motif | Compound | $\lambda_\Gamma$ | $\lambda_{BZ}$ | Synthesized? | Experimental $T_c$ | Calculated $T_c$ | Space group |
|---|---|---|---|---|---|---|---|
| B1 | $C_2B_{13}$ | 0.36 | 0.81[35] | Y[36] | N[37] | 15-30 K[35] | $R\text{-}3m$ |
| B2 | $ZrB_{12}$ | 0.14 | 0.61 | Y[38] | 6.0 K[39] | 6.9 K$^*$ | $Fm\text{-}3m$ |
| B3 | $KB_6$ | 0.13 | 0.56[5] | Y[40] | N[41] | 16.8 K[5] | $Pm\text{-}3m$ |
| B4 | $Mo_2B$ | 0.27 | 0.94 | Y[42] | 6.0 K[43] | 7.0 K$^*$ | $I4/mcm$ |
| B4 | $Ta_2B$ | 0.15 | - | Y[42] | 3 K[44] | - | $I4/mcm$ |
| B5 | WB | 0.34 | 0.56 | Y[45] | 2.0 K[45] | 5.3 K$^*$ | $Cmcm$ |
| B5 | MoB | 0.32 | - | Y[45] | 2.4 K[45] | - | $Cmcm$ |
| B6 | $MgB_2$ | 0.42 | 0.70[7] | Y[49] | 39 K[6] | 22 K[7] | $P6/mmm$ |
| B6 | $TaB_2$ | 0.12 | 0.73[46] | Y[50] | 9 K[53] | 12 K[46] | $P6/mmm$ |
| B6 | $NbB_2$ | 0.10 | 0.71[47] | Y[51] | 9 K[45] | 5.7-19.4 K[47] | $P6/mmm$ |
| B6 | $ScB_2$ | 0.05 | 0.47[48] | Y[52] | 1.5 K[48] | 1.62 K[48] | $P6/mmm$ |
| B7 | $Be_2B$ | 0.34 | 0.51[54] | Y[55] | N[56] | ~ 10 K[54] | $Fm\text{-}3m$ |
| T1 | $Sr(Pt_3B_2)_2$ | 0.82 | 1.59 | Y[57] | 2.7K[57] | 4.1 K$^*$ | $P\text{-}3m1$ |
| T1 | $Ba(Pt_3B_2)_2$ | 0.64 | - | Y[57] | 5 K[57] | - | $P\text{-}3m1$ |
| T2 | $WNbB_2$ | 0.22 | 0.44 | Y[51] | ? | 2.4 K$^*$ | $Amm2$ |
| T2 | $TaMoB_2$ | 0.20 | 0.81 | N | ? | 5 K$^*$ | $Amm2$ |
| T3 | $MoIrB_4$ | 0.10 | 0.48 | N | ? | 4 K$^*$ | $P\text{-}6m2$ |
| T3 | $OsWB_4$ | 0.10 | - | N | ? | - | $P\text{-}6m2$ |
| T4 | $Ca_2Rh_5B_4$ | 0.25 | 0.72 | Y[58] | ? | 4.9 K$^*$ | $Fmmm$ |
| T4 | $Sr_2Rh_5B_4$ | 0.22 | - | Y[58] | ? | - | $Fmmm$ |
| T5 | $TaMo_2B_2$ | 0.93 | 0.81 | Y[59] | ? | 12 K$^*$ | $P4/mbm$ |
| T5 | $NbMo_2B_2$ | 0.57 | 0.76 | Y[60] | ? | - | $P4/mbm$ |
| T6 | $YIr_2B_2$ | 0.38 | 0.97 | Y[61] | 3.3 K[62] | 7.4 K$^*$ | $Fddd$ |

The classified 23 candidates contain 12 binary compounds and 11 ternary compounds. As shown in Fig. 1(c), the elements involved in these compounds before the E filter are evenly distributed in the elemental table and do not show a chemical preference. After applying the EPC filter, the remaining elements are mostly on the left side of the elemental table, showing a preference for group-II elements. Based on the atomic packing, we classify the structures into 7 families for binary compounds and 6 for ternary compounds. The structural information and motifs are shown in Table 1 and Fig. 2, respectively. The boron atoms exhibit diverse atomic packing in these motifs. In Motif B1, B2, and B3, boron atoms pack as $B_{13}$, $B_{12}$, and $B_6$ polyhedra, respectively. In Motif B4 and B5, boron atoms pack as a chain. In Motif B6, boron forms a hexagonal layer. In Motif B7, boron forms a 3D structure with the metal atom. In ternary phases, boron's packing shows a similar motif as the ones in binary phases. In Motif T1, two boron atoms form a dumbbell structure. Motif T2 is essentially the same as Motif B5 except that different metal elements occupy the metal site. Motif T3 is like B6, while the hexagonal boron layer is buckled. In Motif T4, boron and small metal atoms (Rh in the figure) form a square ring. The Motif T5 also has a $B_2$ dumbbell like the T1. In Motif T6, each Ir atom is connected to four surrounding B atoms to form a ring similar to the one in B7.

The screened superconducting phase shows diverse chemical and structural characteristics. The following sections review relevant literature for synthesized superconductors in these structures. We will select a specific compound to conduct full Brillouin zone EPC and $T_c$ calculations for structures that have not been explored. These will validate the reliability of our screening methodology and potentially discover new superconducting materials that have not been previously documented.

3.2 Binary borides

Comparing with the literature, we find that most current binary candidates possess phonon-mediated superconductivity, as shown in Table 1. Except for $MgB_2$, other compounds in the experimental study show $T_c$ lower than 10 K.

$ZrB_{12}$ is a two-gap strongly coupled superconductor with an experimental $T_c$ of ~ 6 K [39,63]. In our calculation, $T_c$ of $ZrB_{12}$ was estimated at 6.92 K by the McMillan-Allen-Dynes (MAD) formula with $\lambda = 0.61$. The low-frequency phonon modes dominated by the Zr atom contribute 0.34 to $\lambda$ as shown in Fig. S12.

Kayhan reported low $T_c$ of ~ 2 K in WB and MoB compounds. The B6 structure is from the MgB$_2$ family, frequently studied [6, 45-53].

Mo$_2$B was recently reported as a weakly coupled superconductor with $T_c$ of ~ 6 K[43]. However, some imaginary phonon modes exist at Γ point, as shown in Fig. S13. It should be noted that the appearance of imaginary frequency modes in the phonon spectrum does not necessarily imply realistic instability. It was suggested that the lack of consideration of anharmonic effects may cause computational instability of $I4/mcm$ Mo$_2$B[64]. The anharmonic effect can stabilize structures, as seen in CaSiO$_3$[65], SrTiO$_3$[66], and Na$_2$TISb[67]. Other effects, such as temperature, charge density wave, vacancies, and structural distortions, can also stabilize the structure. It was suggested the dynamically stable $I4/m$ phase could better explain the Mo$_2$B structure at the harmonic level[68]. However, the structure of the $I4/m$ phase is very similar to the $I4/mcm$ phase, and their XRD data were indistinguishable.

Some predicted superconductors, such as C$_2$B$_{13}$, KB$_6$, and Be$_2$B, were not demonstrated by experiment. The structure of C$_2$B$_{13}$ was similar to B$_4$C, where the B icosahedra were connected by B/C atoms. The C$_2$B$_{13}$ structure was predicted to be a potential superconductor with a $T_c$ of 15-30 K[35]. However, the experimentally synthesized C$_2$B$_{13}$ contains a large number of defects, leading to a semiconducting state that hinders its superconductivity[37].

KB$_6$ was calculated with $T_c$ of 16.8 K[40]. However, KB$_6$ has not yet exhibited superconductivity in experiments. Possible factors include small, mutually insulated grains, inhibitory oxide impurities, adverse electronic localization, low magnetization experiment temperatures, and a disparity between theoretical predictions and practical superconductivity[41]. Be$_2$B was also predicted to be a potential superconductor[69] with $T_c$ of ~ 10 K[54]. Unfortunately, the experimental synthesized Be$_2$B contained significant defects that induced uncertainties on its stoichiometry[56].

3.3 Ternary borides

Table 1 shows several ternary borides as potential superconductors with $\lambda_\Gamma > 0.1$. Notably, among these ternary borides, the T1 family comprising Sr(Pt$_3$B$_2$)$_2$ and Ba(Pt$_3$B$_2$)$_2$ and T6 family (YIr$_2$B$_2$) were confirmed as superconductors by experiment[57,62,70]. The experimentally synthesized Sr-Pt-B phase shows a partial occupancy on Sr sites, resulting in a Sr$_{0.66}$Pt$_3$B$_2$ composition[57]. In our calculation, we used the same structure but a vacancy and a fully occupied Sr site to study this compound, as shown in Fig. S15. This may lead to the imaginary phonon modes shown in Fig.

S15(c). Because the $T_c$ calculations with correct treatment of dynamical instability and partially occupied structure are expensive, we made a rough estimation of $T_c$ for $Sr(Pt_3B_2)_2$ under the assumption that the imaginary phonon modes do not contribute to the EPC, as did in previous high-throughput works[71]. It shows the approximated structure has a similar $T_c$ (4.1 K) compared to experimental data (2.7 K). The full Brillouin-zone EPC calculation results of $YIr_2B_2$ are shown in Fig. S16. Other ternary families have not been investigated for superconductivity by experiment or theory. We select one compound for each family in T2-T5 to perform DFPT calculations. The calculated full Brillouin-zone EPC constant and estimated $T_c$ with the MAD formula are shown in Table 1 and discussed below.

**$WNbB_2$**. $WNbB_2$ shares a similar structure to the orthorhombic WB (space group *Cmcm*). The B atoms in $WNbB_2$ exist as a one-dimensional B atom chain see Fig. 3(b). As in Fig. 3(a), a van-Hove singularity is near the Fermi level, mainly dominated by Nb-*4d* orbital and W-*5d* orbital. On the other hand, due to the significant mass differences between B and Nb/W, the phonon modes are entirely decoupled. Nb and W contribute to the EPC of low phonon frequency modes, while B contributes to the EPC of high phonon frequency modes. Fig. 3(b) shows the phonon mode with a frequency of 13.35 THz at Γ point, which possesses large phonon linewidth, contributing major EPC strength in high-frequency phonons. The $T_c$ of $WNbB_2$ is ~ 2 K, smaller than WB (5.3K). The main reason for the difference in $T_c$ between $WNbB_2$ and WB is that the EPC contributed by low-frequency phonons in WB is higher than in $WNbB_2$. Compared with WB, introducing the Nb element in $WNbB_2$ increases the frequency of some low-frequency optical phonons, suppresses the softening of phonons, and weakens EPC.

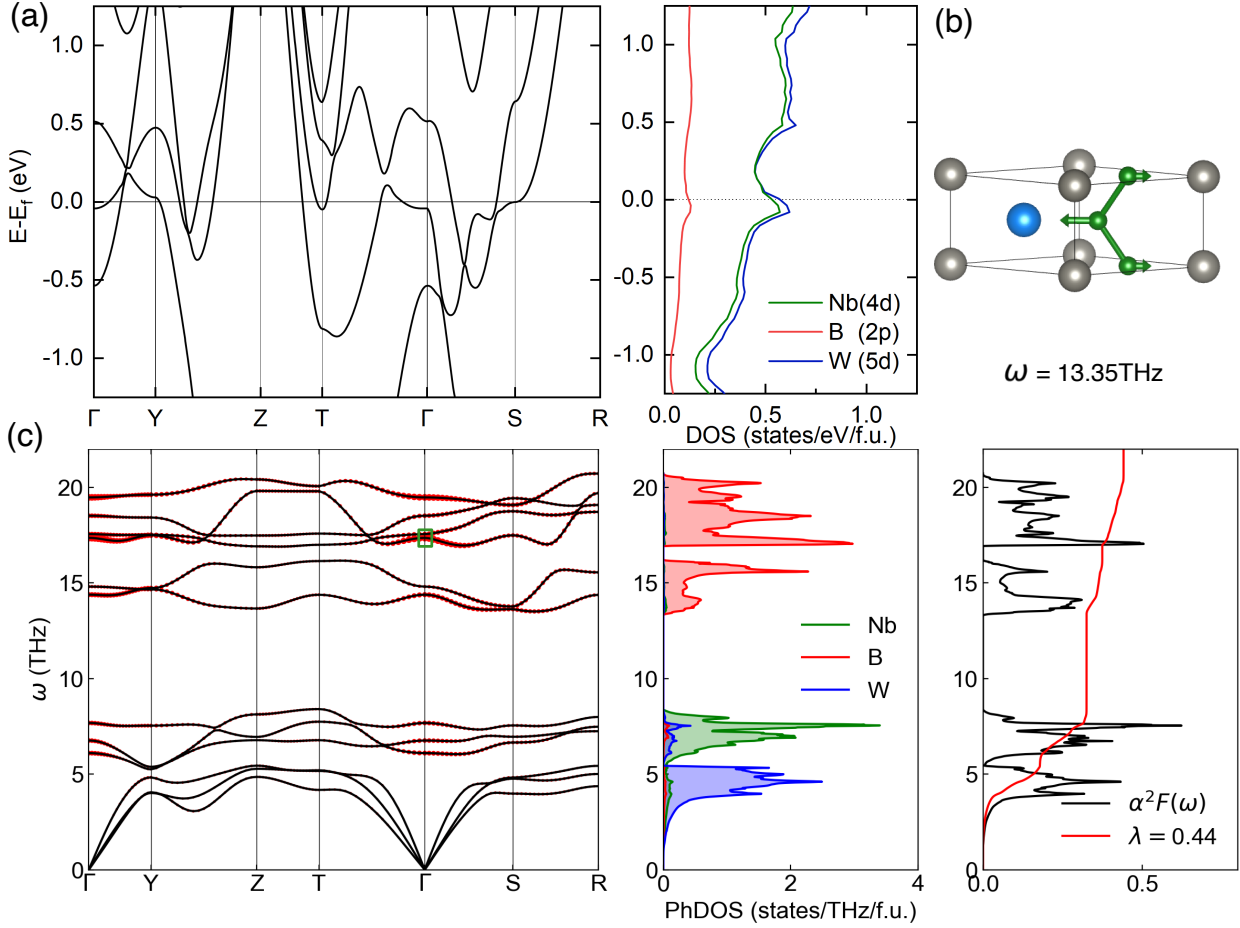

**Fig. 3** (a) Band structure and projected density of states of WNbB$_2$. (b) Structure of WNbB$_2$. The arrow represents the eigenvector of phonon mode at the Γ point with a frequency of 13.35 THz, noted in (c). Blue is W, gray is Nb, and green is B. (c) Phonon spectrum, projected phonon density of state, and Eliashberg spectral function of WNbB$_2$. The red bands on the phonon spectrum indicate the phonon linewidth.

**MoIrB$_4$.** The band structure and phonon spectrum of MoIrB$_4$ are shown in Fig. 4. Projected electron density of states indicates that Ir-$5d$ orbital dominates the $N(E_F)$, and Ir contributes significantly to EPC. The calculated $T_c$ of MoIrB$_4$ is 4K. MoIrB$_4$ has a ReB$_2$-type structure and exhibits layered characteristics. It shows some analogy to MgB$_2$, in which the B atoms form interlayer hexagonal rings between metal layers. The distinction between MoIrB$_4$ and MgB$_2$ lies in the fact that the B hexagonal rings in MoIrB$_4$ are not coplanar but adopt a buckled three-dimensional conformation. This buckled structure may originate from the metal atoms above the B rings instead of at the ring centers, which compress the B rings and lead to corrugation deformation. The EPC strength in MgB$_2$ is mainly contributed by the in-plane vibration mode of B atoms[72]. However, the situation is different in MoIrB$_4$. The phonon mode with the largest linewidth among high-frequency phonons corresponds to the out-of-plane vibration of B atoms, as shown in Fig. 4(b). Moreover, Figure 4(c) shows the EPC in MoIrB$_4$ is mainly from Ir atoms with ~60% contribution and only ~20% from B atoms. While the synthesis of stoichiometric MoIrB$_4$ was not reported, a few ReB$_2$-type ternary borides were briefly reported in the 1970s[73,74]. More careful experimental investigations on these systems might reveal some superconductors. The Mo-Ir-B system has another ternary Mo$_2$IrB$_2$ phase, which was synthesized experimentally[75]. Vandenberg et al.[76] reported $T_c$=3.6 K for the Mo$_2$IrB$_2$ compound. In our screening calculations, Mo$_2$IrB$_2$ did not pass our E-filter because the zone-center modes made little contribution to the superconductivity in this compound.

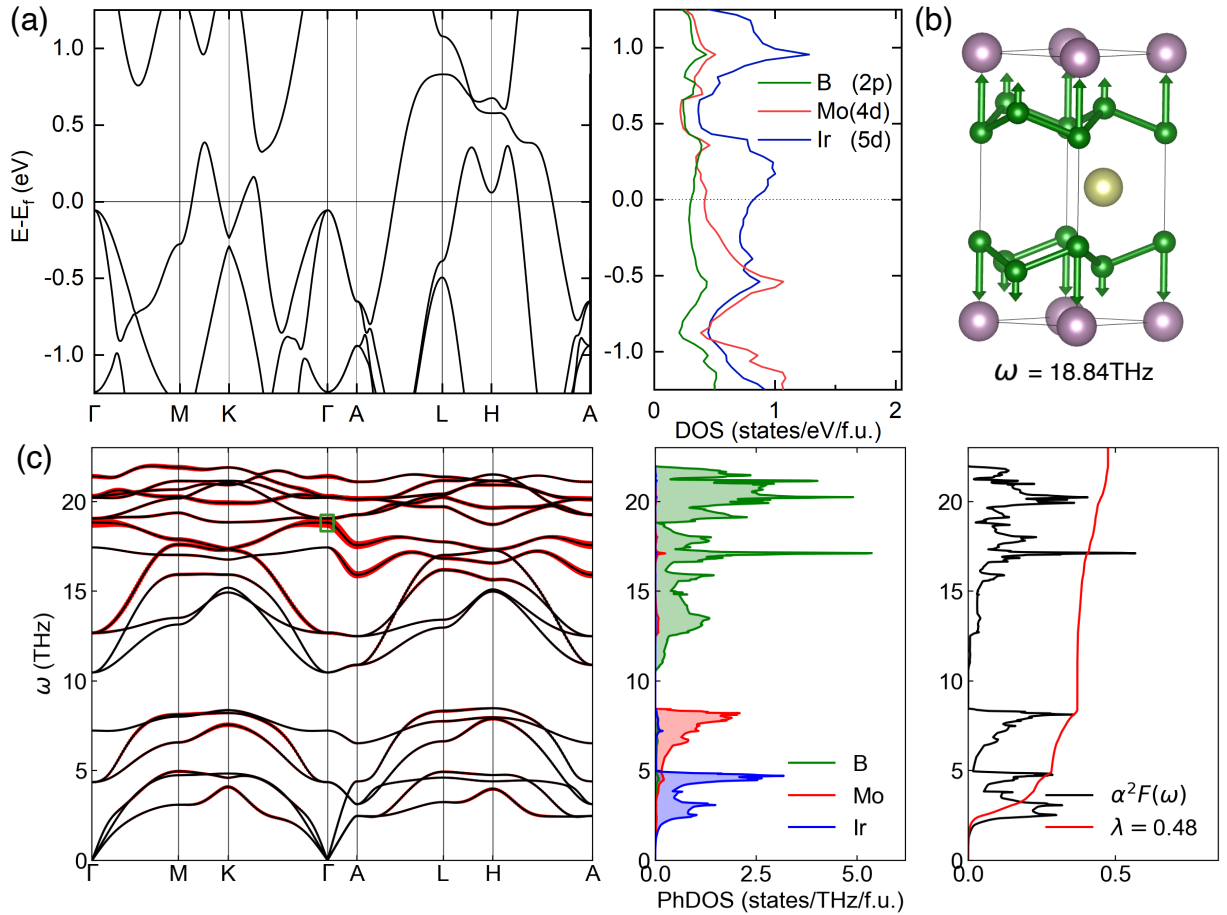

**Fig. 4** (a) Band structure and projected density of states of MoIrB$_4$. (b) Structure of MoIrB$_4$. The arrow represents the eigenvector of phonon mode at Γ point with a frequency of 18.84 THz, noted in (c). Purple is Mo, yellow is Ir, and green is B. (c) Phonon spectrum, projected phonon density of state and Eliashberg spectrum of MoIrB$_4$. The red bands on the phonon spectrum indicate the phonon linewidth.

**Ca$_2$Rh$_5$B$_4$.** Rh atom in Ca$_2$Rh$_5$B$_4$ is coplanar with the four surrounding B atoms with Rh-B bond length ~2.16Å, forming a series of RhB$_4$ clusters. These clusters are connected by sharing Rh atoms to form a framework. Ca atoms are embedded in this matrix framework. This structure was first synthesized in 1983[58]. The band structure and phonon spectrum are shown in Fig. 5(a) and (c). Among B-dominated high-frequency phonons, the phonon mode with the largest phonon linewidth is the 26$^{th}$ mode at Γ point ($\omega$= 13.19 THz), which is the breathing vibration of B atoms in the RhB$_4$ cluster, shown in Fig. 5(b). The DFPT result indicates that Ca$_2$Rh$_5$B$_4$ is a potential superconductor with $T_c$ = 4.9 K and $\lambda$=0.72.

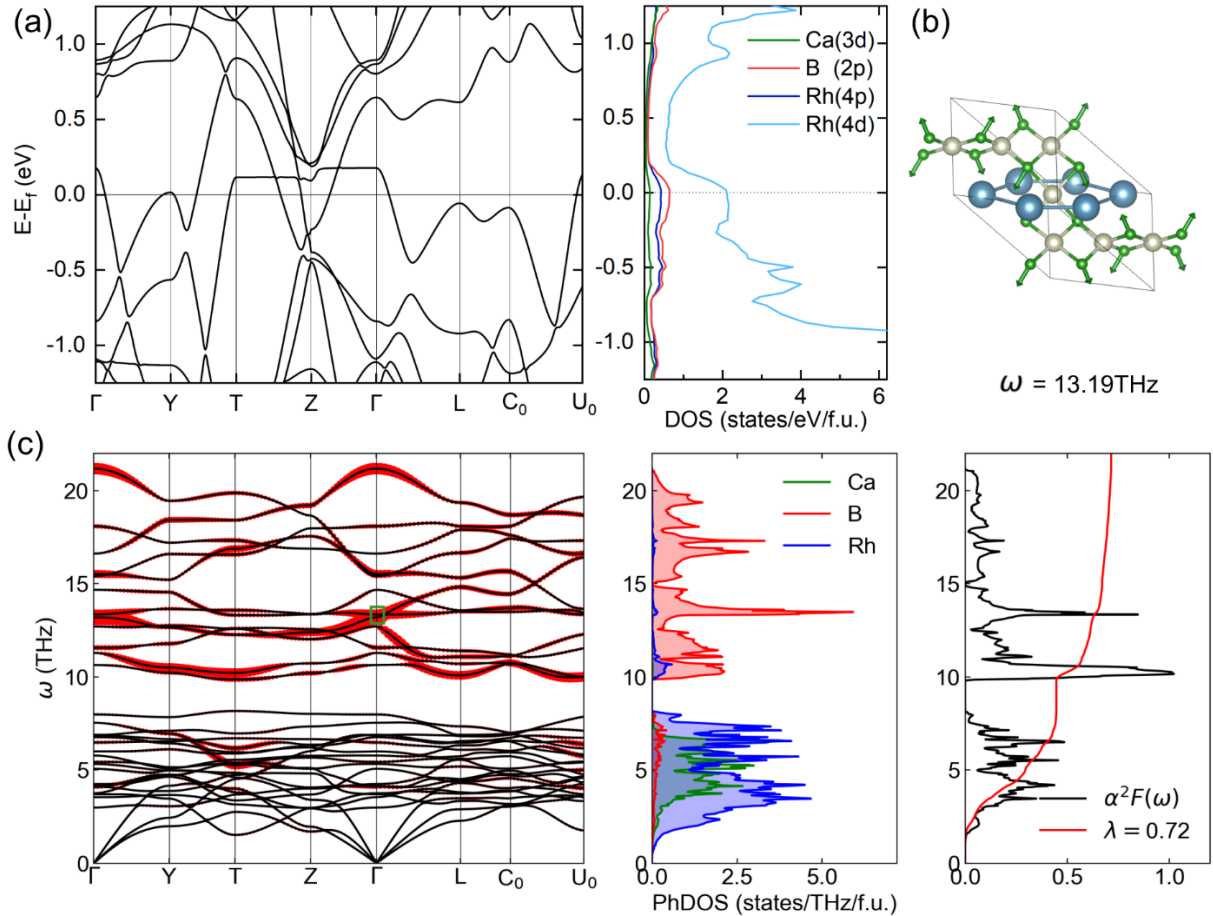

**Fig. 5** (a) Band structure and projected density of states of Ca$_2$Rh$_5$B$_4$. (b) Structure of Ca$_2$Rh$_5$B$_4$. The arrow represents the eigenvector of the phonon mode at Γ point with a frequency of 13.19 THz, noted in (c). Blue is Ca, yellow is Rh and green is B. (c) Phonon spectrum, projected phonon density of state, and Eliashberg spectrum of Ca$_2$Rh$_5$B$_4$. The red bands on the phonon spectrum indicate the strength of EPC.

**TaMo$_2$B$_2$**. While Kuz'ma et al. [59] reported a solid solution phase for the (Ta$_{1-x}$Mo$_x$)$_3$B$_2$, we consider the ordered structure for the stoichiometric TaMo$_2$B$_2$ compound to investigate its superconductivity. The B$_2$ dumbbell in TaMo$_2$B$_2$ has a B-B bond length of ~ 1.89 Å. Although the contribution of B-dominated high-frequency phonon modes to EPC strength is weak, phonon mode dominated by heavy elements in the low-frequency region contributes to large EPC. $T_c$ of TaMo$_2$B$_2$ is ~12 K and $\lambda$=0.81. One of the modes that contributes significantly to the EPC in the low-frequency region is the in-plane stretching vibration, primarily involving Mo atoms and, to a lesser extent, B atoms, as shown in Fig. 6(b). On the other hand, high electron density of states benefits strong EPC. The van Hove singularity below the Fermi level may regulate the $E_f$ by hole doping or induce high-frequency phonon modes to soften by element doping, which provides a promising way to improve further the superconducting temperature of TaMo$_2$B$_2$.

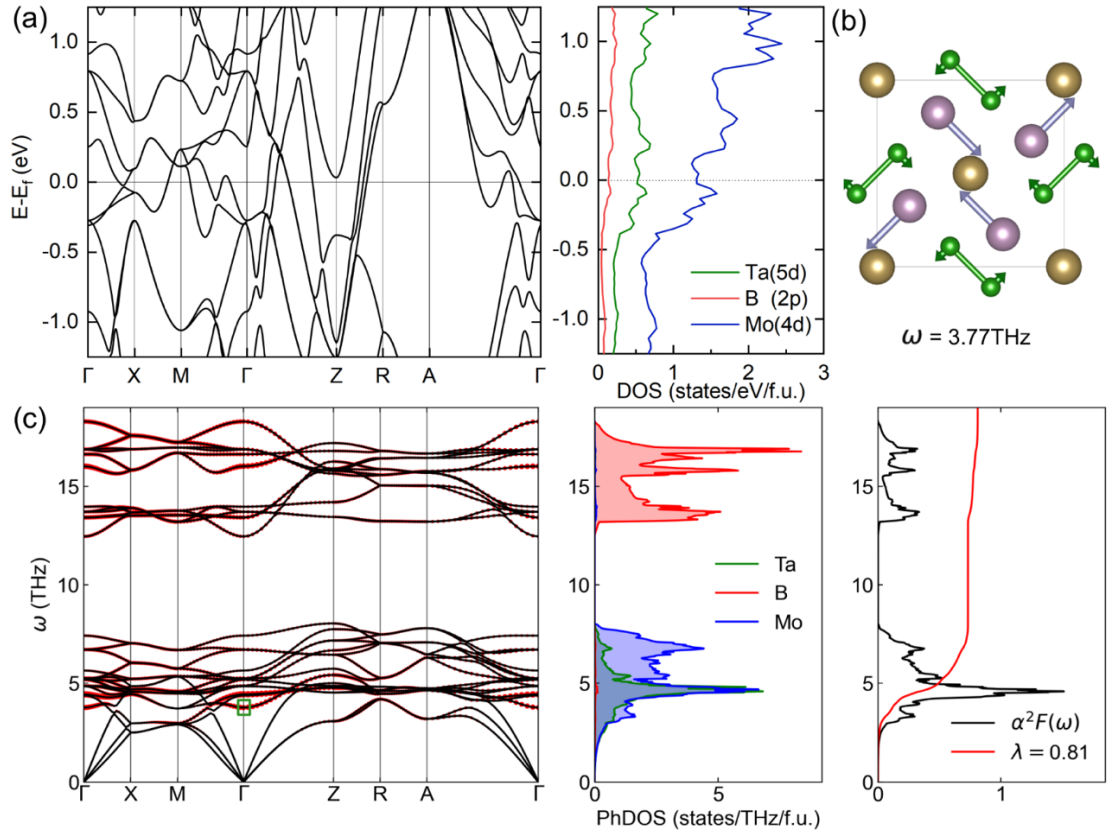

**Fig. 6** (a) Band structure and projected density of states of TaMo$_2$B$_2$. (b) Structure of TaMo$_2$B$_2$. The arrow represents the eigenvector of phonon mode at Γ point with a frequency of 3.8 THz, noted in (c). Purple is Mo, yellow is Ta, green is B. (c) Phonon spectrum, projected phonon density of state, and Eliashberg spectral function of TaMo$_2$B$_2$. The red bands on the phonon spectrum indicate the phonon linewidth.

We have shown that large-scale screening with fast EPC calculations can efficiently suggest potential conventional superconductors for future experimental investigations. By comparing the EPC constant between full Brillouin zone DFPT calculations and the zone-center EPC strength in Table 1, we found a rather positive correlation, as shown in Fig. S17. This correlation is similar to the one in hydrides identified previously.[12] However, in contrast to the unique hydrogen network found in high-pressure superconducting hydrides, the patterns of superconductivity in borides appear to be more diverse. The low-frequency phonon bands involving the metal and boron atoms contribute significantly to EPC in many boride structures. We also notice the limitation of the scheme: the EPC screening implemented in the E-filter may overlook systems where the strong EPC is concentrated away from the Brillouin zone center, for example, in the $Mo_2IrB_2$ compound with experimental $T_c$=3.6 K [76]. Such "false negative" cases are hard to avoid with our simplified EPC estimation but do not affect the accuracy of the predicted superconducting phases. Our calculations on the superconductivity are all based on the ordered structure, while experiments often resulted in solid solution phases. A careful design of synthetic paths can be crucial to realizing the predicted compounds.

**IV. Conclusion**

In summary, by applying a series of filters to extract binary and ternary boride candidates from the crystal structure database, we identified 23 promising candidates as potential superconductors. Further DFPT calculations revealed their EPC and superconducting temperature. While superconductivity in binary borides has been confirmed through experiments and computations, the superconductivity in ternary borides remains relatively unexplored. Notably, $TaMo_2B_2$ exhibits the highest superconducting temperature at approximately 12 K. Since most ternary compounds have been synthesized, the predicted superconductivity can be validated through future experiments. This study highlights the potential of utilizing high-throughput computational screening to discover new superconducting materials. Additionally, it demonstrates that, apart from the well-studied $MgB_2$ structure, diverse structural motifs in borides can lead to superconductivity. These motifs could be the foundation for future high-throughput screening involving elemental substitution to discover new compounds.

**Supporting Information:** Additional figures for zone-center EPC strength, phonon dispersion, phonon density of state, Eliashberg spectrum and convergence test for the $T_c$ calculations (PDF).

**Acknowledgement**

The work at Xiamen University was supported by the Natural Science Foundation of Xiamen (Grant No. 3502Z202371007) and the Fundamental Research Funds for the Central Universities (Grant No. 20720230014). V.A. was supported by the U.S. Department of Energy, Office of Basic Energy Sciences, Division of Materials Sciences and Engineering. Ames National Laboratory is operated for the U.S. Department of Energy by Iowa State University under Contract No. DE-AC02-07CH11358. K.-M.H. acknowledges support from National Science Foundation Award No. DMR2132666. Shaorong Fang and Tianfu Wu from the Information and Network Center of Xiamen University are acknowledged for their help with Graphics Processing Unit (GPU) computing.